\documentclass[a4paper,11pt]{article}
\usepackage{pos}
\usepackage{natbib}
\usepackage{aas_macros}

\title{Cosmic Ray Transport and Gamma-Ray Signatures in the Interstellar Medium}

\author*[a]{Lucas Barreto-Mota}
\author[a]{Elisabete M. de Gouveia Dal Pino}
\author[b]{Siyao Xu}
\author[c]{Alexandre Lazarian}
\author[d,e,f]{Rafael Alves-Batista}
\author[e,f]{Gaetano Di Marco}
\author[a]{Stela Adduci Faria}

\affiliation[a]{Instituto de Astronomia, Geofísica e Ciências Atmosféricas, Universidade de São Paulo\\
R. do Matão, 1226 - Butantã, São Paulo - SP, 05508-090
Brazil}

\affiliation[b]{Department of Physics, University of Florida\\ 
2001 Museum Rd, Gainesville, FL 32611 , USA}

\affiliation[c]{University of Wisconsin, Department of Astronomy\\
475 Charter Street, Madison, WI 53706, USA}

\affiliation[d]{Institut d’Astrophysique de Paris (IAP), Sorbonne Universit´e\\
CNRS, UMR 7095, 98 bis bd Arago, 75014 Paris, France}

\affiliation[e]{Departamento de Física Teórica, M-15, Universidad Autónoma de Madrid, E-28049 Madrid, Spain}

\affiliation[f]{Instituto de Física Téorica UAM/CSIC\\
Calle Nicolás Cabrera 13-15, Cantoblanco, 28049 Madrid, Spain}

\emailAdd{lucas.barreto.santos@usp.br}

\abstract{The interaction of cosmic rays (CRs) with magnetic fields and the interstelar medium (ISM) leads to the production of nonthermal radiation. Although this has been a topic of study for many years, it still poses many challenges to the understanding of these processes. In this work we present a short review of recent advances in the understanding of CR propagation in magnetohydrodynamical (MHD) turbulence, in particular the process of mirror diffusion, and how it can help explain recent observational constraints for CR diffusion away from sources. We also present preliminary results from Monte Carlo simulations of CR cascading and propagation within  a young massive stellar cluster (YMSC), aimed at probing the origin of very-high-energy (VHE) emission from these sources.}

\ConferenceLogo{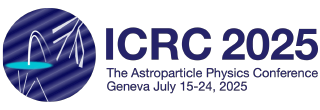}

\FullConference{39th International Cosmic Ray Conference (ICRC2025)\\
 15–24 July 2025\\
Geneva, Switzerland\\}

\begin{document}
\maketitle

\section{Introduction}

The interaction of cosmic rays (CRs) with interstellar magnetic fields and gas gives rise to various non-thermal emission processes. The observed CR energy spectrum follows a power law, featuring two major breaks: the knee and the ankle. The ankle occurs at ultra-high energies and is likely attributable to the GZK effect. In contrast, the knee, which appears at lower energies, is commonly associated with the transition from Galactic to extragalactic cosmic ray sources.


This transition likely occurs around a few hundred PeV, an energy regime that confines the number of viable Galactic accelerators—known as Pevatrons—to only a handful of candidate sources.

Recent observations provide crucial constraints on particle behavior in these environments. H.E.S.S. data suggests that the diffusion coefficient for CR electrons in the vicinity of the Vela X pulsar wind nebula is $\lesssim10^{28}$ cm²/s at 10 TeV within the inner tens of parsecs \cite{2018ApJ...866..143H}. Similar suppressed diffusion has been inferred from HAWC observations of the Geminga and PSR B0656+14 pulsar wind nebulae \citep{2017Sci...358..911A}.

Even more compelling is the recent LHAASO detection of very high-energy (VHE) emission extending to hundreds of TeV, which is compatible with the presence of multiple PeVatron sources within our Galaxy \citep{2021Natur.594...33C}. The potential sources linked to these observations include molecular cloud-supernova remnant (MC-SNR) interaction regions \citep[e.g.,][]{PhysRevLett.108.051105,PhysRevLett.109.061101,2013Sci...339..807A}, pulsar wind nebula halos (TeV halos) \citep{PhysRevD.104.123017,yan2023origin,2024NatAs.tmp...54Y}, and Young Star Clusters (YSCs) \citep{2021MNRAS.504.6096M,2021NatAs...5..465A}.




Classically, the diffusion of cosmic rays (CRs) is described by quasi-linear theory (QLT) \citep{1966ApJ...146..480J}. However, standard QLT struggles to reproduce the severely suppressed diffusion coefficients inferred from observations. It also fails to resolve the so-called $90^\circ$ scattering problem, which concerns the change in propagation direction for particles with pitch angles perpendicular to the magnetic field.

Various mechanisms have been proposed to explain these phenomena \citep[e.g.,][]{2022FrASS...922100F}. In this work, we focus on mirror diffusion \citep{2021ApJ...923...53L}, a non-resonant mechanism that can self-consistently account for both suppressed diffusion and efficient $90^\circ$ scattering.


Classically, particles interacting with magnetic mirrors were considered to be trapped within magnetic bottles, a process not thought to contribute to spatial diffusion. However, \cite{2021ApJ...923...53L} demonstrated that when superdiffusion of magnetic field lines is present, particles scattering off magnetic mirrors do not simply retrace their paths. Instead, their trajectories are randomized by the meandering of the field lines, preventing them from returning to their previous mirroring point \citep[see also][]{2013Natur.497..466E,2022MNRAS.512.2111H,2023ApJ...959L...8Z}. Given that MHD turbulence naturally generates magnetic compressions (which act as mirrors) and simultaneously causes the superdiffusive spreading of field lines, mirror diffusion provides a self-consistent mechanism for explaining both the change in CR propagation direction and the suppression of diffusion parallel to the mean magnetic field.

Recently, combining 3D MHD simulations of turbulent star-forming regions \citep{barreto2021} with test particle simulations to analyze CR diffusion,  \cite{2025ApJ...988..269B} demonstrated that mirror diffusion can produce diffusion coefficients as low as $\lesssim10^{27}$ cm²/s in the ISM, even under conditions of suppressed resonant scattering. Building on this foundation, our work extends the exploration of suppressed diffusion near Galactic sources by focusing on its impact on the production of secondary particles, simulated using CRPropa \citep{Alves_Batista_2022}.

\section{Cosmic Ray propagation simulations inside 
YMSCs}



To investigate the implications of mirror diffusion for the production of secondary particles, we conducted numerical simulations using the  CRPropa  framework \citep{Alves_Batista_2022}. CRPropa is a publicly available Monte Carlo code designed to simulate the propagation of cosmic rays, gamma-rays, and neutrinos from their sources, through various photon backgrounds and magnetic fields, to an observer.

We embedded our CRPropa simulations within a realistic astrophysical environment provided by the 3D magnetohydrodynamic (MHD) simulations of Young Massive Star Clusters (YMSCs) from \cite{2024A&A...690A..94P}. Specifically, we adopted their intermediate model, which simulates the formation of a cluster with characteristics analogous to W43. This model self-consistently includes the formation of thousands of stars, dust temperature, and UV radiation field feedback, alongside the standard MHD quantities such as gas density and magnetic field strength. This provides a highly realistic background for studying cosmic-ray propagation and interaction.


A current limitation of CRPropa is that it only accepts isotropic photon fields for interaction calculations. To model the radiation environment within the YMSC as a first approximation, we constructed a composite isotropic photon field from three components. For the interstellar radiation field (ISRF), we adopted the standard model provided in the CRPropa examples, based on \cite{2017ApJ...846...67P}. The dust emission was approximated as a blackbody spectrum using the information provided by the MHD simulation. Finally, the stellar contribution was calculated by summing the emission from all stars in the cluster, with each star treated as a blackbody emitter. The radius and temperature of each star were scaled with its mass according to the relationships:

\begin{align}
    R_{*}=R_\odot \times \Big(\frac{M_{*}}{M_\odot}\Big)^{0.8}\\
    T_{*}=5778K \times \Big(\frac{M_{*}}{M_\odot}\Big)^{0.5},
\end{align}

\noindent where $R_\odot$ and $M_\odot$ are the radius and mass of the Sun. 

\begin{figure*}[ht]
\begin{center}
    \includegraphics[width=0.5\columnwidth,angle=0]{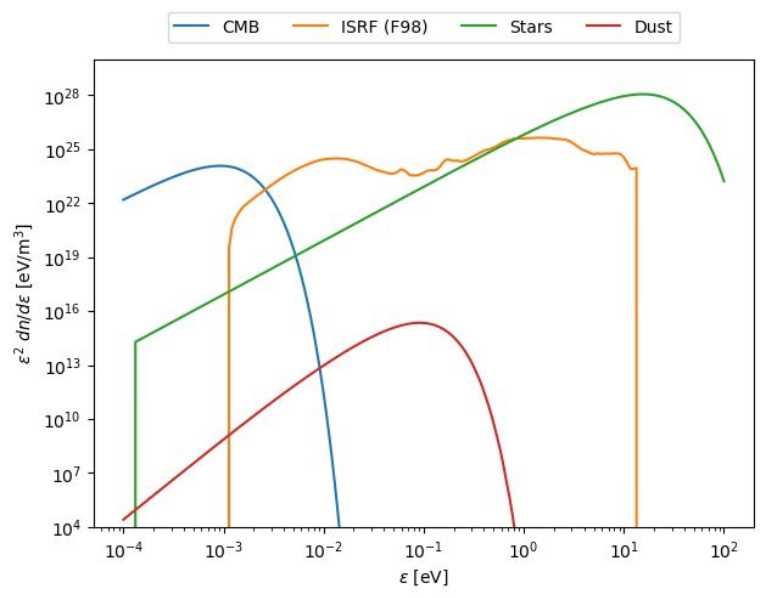}
    
    \caption{Photon energy density for the different background photon fields. Blue line shows the contribution for the CMB, orange for the background interstellar radiation field, green for the stars 
    \textbf{inside} the YMSC, and red for the dust present in the cluster.
    } 
    \label{fig:background_radiation}
\end{center}
\end{figure*}

The photon density contribution of each background photon field is shown in Figure \ref{fig:background_radiation} where the cosmic microwave background is also included.

{}

\section{Results}

\begin{figure*}[ht]
\begin{center}
    \includegraphics[width=0.6\columnwidth,angle=0]{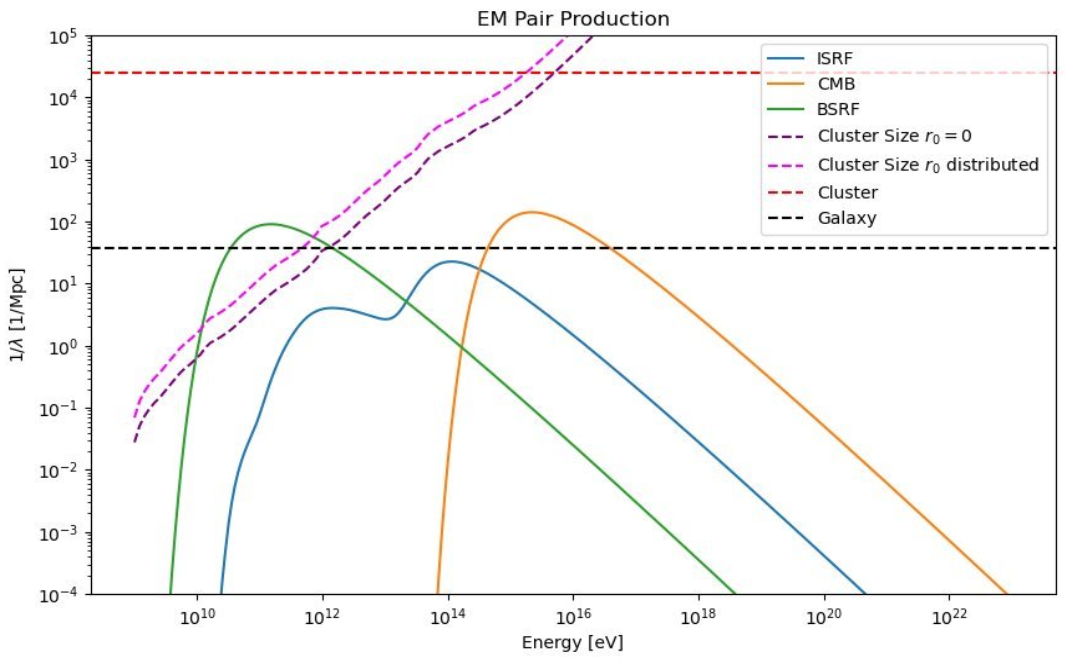}
    \includegraphics[width=0.6\columnwidth,angle=0]{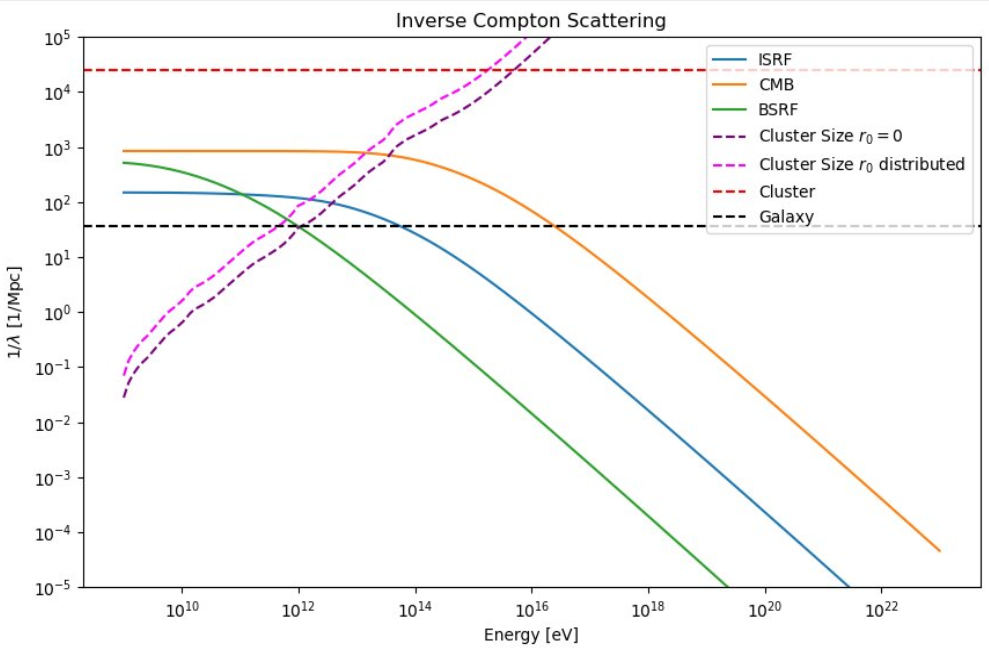}
    
    \caption{The inverse mean free path (MFP) for two interaction processes. Top: Pair production. Bottom: Inverse Compton scattering. Solid colored lines represent the contribution from different radiation fields: interstellar radiation field (ISRF, blue), cosmic microwave background (CMB, orange), and stellar blackbody emission from the cluster (green). Dashed lines indicate key spatial scales: the 40 pc cluster size (red), the Galactic scale (black), and the effective confinement scales from mirror diffusion for a central source (purple) and a uniform source distribution (pink).
    } 
    \label{fig:inverse_mfp_em_ic}
\end{center}
\end{figure*}{}

\begin{figure*}[ht]
\begin{center}
    \includegraphics[width=0.6\columnwidth,angle=0]{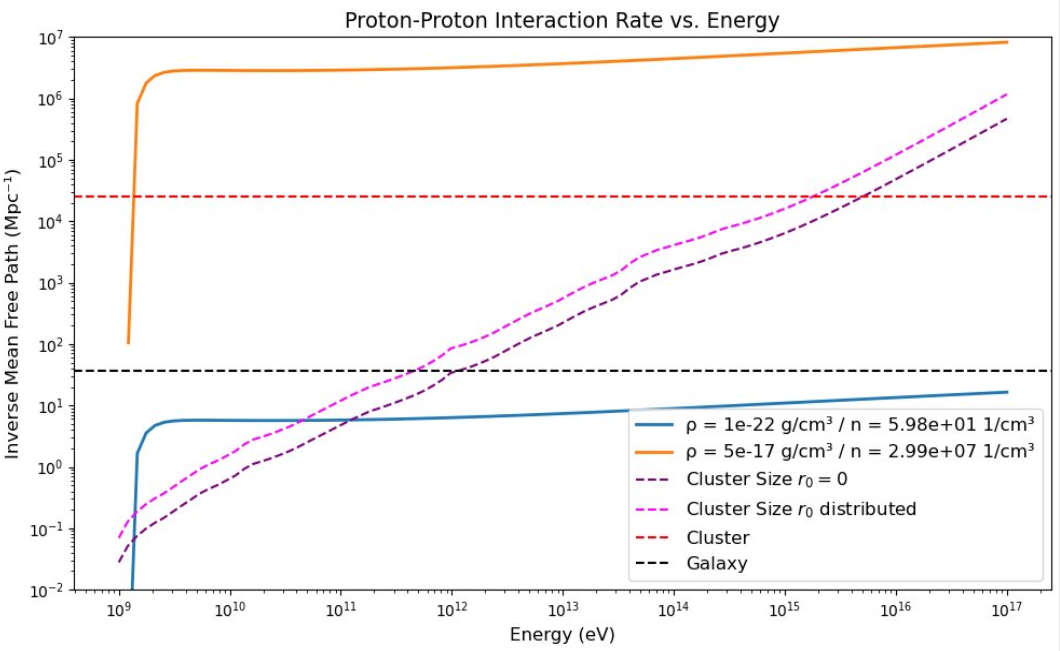}
    
    \caption{Inverse mean free path (MFP) for proton-proton (pp) interactions, calculated following the parameterization from \cite{2006PhRvD..74c4018K}. The blue and orange lines indicate the interaction MFP for the lowest and highest gas densities found in any cell of the magnetohydrodynamic simulation, respectively. These values represent the range of expected interaction lengths within the young massive star cluster environment.
    Dashed lines indicate key spatial scales: the 40 pc cluster size (red), the Galactic scale (black), and the effective confinement scales from mirror diffusion for a central source (purple) and a uniform source distribution (pink).
    } 
    \label{fig:inverse_mfp_pp}
\end{center}
\end{figure*}{}






An evaluation of the relevant interaction processes within the system identifies three dominant mechanisms: electromagnetic pair production, inverse Compton scattering, and proton-proton interactions. The resulting interaction mean free paths (MFPs) for these processes are presented in Figures \ref{fig:inverse_mfp_em_ic} and \ref{fig:inverse_mfp_pp}.

Figure \ref{fig:inverse_mfp_em_ic} shows the inverse MFP for pair production (top panel) and inverse Compton scattering (bottom panel). The solid lines represent the contributions from the three dominant photon fields: the blackbody stellar radiation field (BSRF) from cluster stars (green), the interstellar radiation field (ISRF; blue), and the cosmic microwave background (CMB; orange). For context, the dashed lines indicate key spatial scales: the physical size of the cluster (40 pc; red) and the Galactic scale (black).

We also plot the estimated average path length traveled by a cosmic ray within the cluster under a mirror diffusion regime \citep{2021ApJ...923...53L}. The purple dashed line shows this scale for a point source located at the center of a spherical cluster:

\begin{align}
\langle l_{\mathrm{center}} \rangle = \frac{c R^2}{15 D(E)},
\end{align}

\noindent where $R$ is the cluster radius, $c$ is the speed of light, and $D(E)$ is the energy-dependent diffusion coefficient. The pink dashed line shows the equivalent scale for a uniform distribution of sources throughout the cluster volume:

\begin{align}
\langle l_{\mathrm{distributed}} \rangle = \frac{c R^2}{6 D(E)}.
\end{align}

These diffusion lengths provide a critical benchmark for assessing whether particles are effectively confined within the cluster long enough to undergo significant interactions.



Figure \ref{fig:inverse_mfp_pp} presents the inverse mean free path for proton-proton (pp) interactions. Analogous to Figure \ref{fig:inverse_mfp_em_ic}, the blue and orange lines represent the interaction MFP for the lowest and highest gas densities found in any cell of the simulation, respectively. These curves were calculated following the parameterization from \cite{2006PhRvD..74c4018K}.

These preliminary results lead to a key finding: without the significantly increased confinement due to suppressed mirror diffusion, the characteristic size of the system (40 pc) is too small for most interactions to occur. Consequently, the production of secondary particles would be drastically reduced, highlighting the critical role of slow diffusion in making these environments visible secondary emitters.

\begin{figure*}[ht]
\begin{center}
    \includegraphics[width=0.6\columnwidth,angle=0]{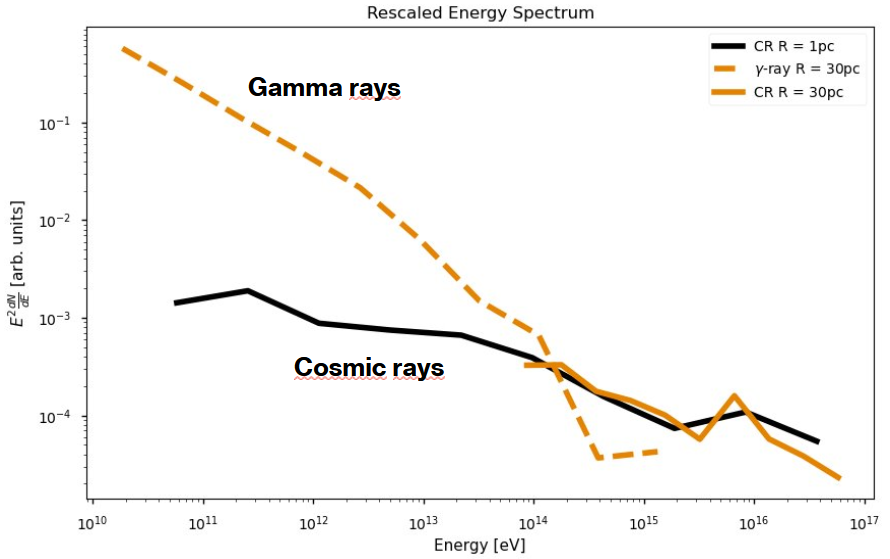}
    
    \caption{Total spectra of $\gamma$-rays and cosmic rays (CRs) from the cascading simulation in arbitrary units. The black solid line represents the injected CR spectrum, measured at a distance of 1 pc from the central source. The orange solid line and orange dashed line show the resulting CR and $\gamma$-ray spectra, respectively, observed at a distance of 30 pc from the source.
    } 
    \label{fig:obs_spectra}
\end{center}
\end{figure*}{}



Figure \ref{fig:obs_spectra} presents the resulting spectra from a preliminary CR propagation simulation considering a single central CR source and two observer configurations. We injected 1000 particles with energies between 10 GeV and 100 PeV. The black solid line shows the injected spectrum of cosmic rays, measured 1 pc from the central CR source. The orange solid line shows the spectrum of cosmic rays collected within a sphere of radius 30 pc, indicating that only the highest-energy particles propagate to this distance. The corresponding spectrum of observed $\gamma$-rays at 30 pc is shown by the orange dashed line.

While these preliminary results successfully reproduce the characteristic energy range and power-law dependence of observed $\gamma$-rays and cosmic rays, it is important to note the limitations of this initial study. The current simulation
relies on very low statistics.
Furthermore, additional background contributions, such as bremsstrahlung emission from HII regions, remain to be evaluated in forthcoming work for a complete picture.

\section{Conclusions}


In conclusion, mirror diffusion provides a compelling mechanism to explain key observational puzzles and resolve long-standing theoretical problems in cosmic-ray propagation. Given the ubiquity of MHD turbulence throughout the universe, this non-resonant diffusion process is likely to operate in a wide range of astrophysical environments, with significant implications not only for particle transport but also for acceleration mechanisms.

While the current CR propagation modeling remains ongoing, the preliminary results are promising and align with theoretical expectations. A complete analysis of these findings, incorporating higher statistics and additional physical processes, will be presented in a forthcoming publication.


\bibliographystyle{JHEP}
\bibliography{refs}

\providecommand{\href}[2]{#2}\begingroup\raggedright\begin{thebibliography}{10}

\bibitem{2018ApJ...866..143H}
Z.-Q.~{Huang}, K.~{Fang}, R.-Y.~{Liu} and X.-Y.~{Wang}, \emph{{Inefficient Cosmic-Ray Diffusion around Vela X: Constraints from H.E.S.S. Observations of Very High-energy Electrons}}, \href{https://doi.org/10.3847/1538-4357/aadfed}{\emph{\apj} {\bfseries 866} (2018) 143} [\href{https://arxiv.org/abs/1807.04182}{{\ttfamily 1807.04182}}].

\bibitem{2017Sci...358..911A}
A.U.~{Abeysekara}, A.~{Albert}, R.~{Alfaro}, C.~{Alvarez}, J.D.~{{\'A}lvarez}, R.~{Arceo} et~al., \emph{{Extended gamma-ray sources around pulsars constrain the origin of the positron flux at Earth}}, \href{https://doi.org/10.1126/science.aan4880}{\emph{Science} {\bfseries 358} (2017) 911} [\href{https://arxiv.org/abs/1711.06223}{{\ttfamily 1711.06223}}].

\bibitem{2021Natur.594...33C}
Z.~{Cao}, F.A.~{Aharonian}, Q.~{An}, L.X.~{Axikegu}, Bai, Y.X.~{Bai}, Y.W.~{Bao} et~al., \emph{{Ultrahigh-energy photons up to 1.4 petaelectronvolts from 12 {\ensuremath{\gamma}}-ray Galactic sources}}, \href{https://doi.org/10.1038/s41586-021-03498-z}{\emph{\nat} {\bfseries 594} (2021) 33}.

\bibitem{PhysRevLett.108.051105}
A.~Neronov, D.V.~Semikoz and A.M.~Taylor, \emph{Low-energy break in the spectrum of galactic cosmic rays}, \href{https://doi.org/10.1103/PhysRevLett.108.051105}{\emph{Phys. Rev. Lett.} {\bfseries 108} (2012) 051105}.

\bibitem{PhysRevLett.109.061101}
P.~Blasi, E.~Amato and P.D.~Serpico, \emph{Spectral breaks as a signature of cosmic ray induced turbulence in the galaxy}, \href{https://doi.org/10.1103/PhysRevLett.109.061101}{\emph{Phys. Rev. Lett.} {\bfseries 109} (2012) 061101}.

\bibitem{2013Sci...339..807A}
M.~{Ackermann}, M.~{Ajello}, A.~{Allafort}, L.~{Baldini}, J.~{Ballet}, G.~{Barbiellini} et~al., \emph{{Detection of the Characteristic Pion-Decay Signature in Supernova Remnants}}, \href{https://doi.org/10.1126/science.1231160}{\emph{Science} {\bfseries 339} (2013) 807} [\href{https://arxiv.org/abs/1302.3307}{{\ttfamily 1302.3307}}].

\bibitem{PhysRevD.104.123017}
S.~Recchia, M.~Di~Mauro, F.A.~Aharonian, L.~Orusa, F.~Donato, S.~Gabici et~al., \emph{Do the geminga, monogem and psr j0622+3749 $\ensuremath{\gamma}$-ray halos imply slow diffusion around pulsars?}, \href{https://doi.org/10.1103/PhysRevD.104.123017}{\emph{Phys. Rev. D} {\bfseries 104} (2021) 123017}.

\bibitem{yan2023origin}
K.~Yan, R.-Y.~Liu, R.~Zhang, C.-M.~Li, Q.~Yuan and X.-Y.~Wang, \emph{On the origin of galactic diffuse tev-pev emission: Insight from lhaaso and icecube},  2023.

\bibitem{2024NatAs.tmp...54Y}
K.~{Yan}, R.-Y.~{Liu}, R.~{Zhang}, C.-M.~{Li}, Q.~{Yuan} and X.-Y.~{Wang}, \emph{{Insights from LHAASO and IceCube into the origin of the Galactic diffuse teraelectronvolt-petaelectronvolt emission}}, \href{https://doi.org/10.1038/s41550-024-02221-y}{\emph{Nature Astronomy} (2024) } [\href{https://arxiv.org/abs/2307.12363}{{\ttfamily 2307.12363}}].

\bibitem{2021MNRAS.504.6096M}
G.~{Morlino}, P.~{Blasi}, E.~{Peretti} and P.~{Cristofari}, \emph{{Particle acceleration in winds of star clusters}}, \href{https://doi.org/10.1093/mnras/stab690}{\emph{\mnras} {\bfseries 504} (2021) 6096} [\href{https://arxiv.org/abs/2102.09217}{{\ttfamily 2102.09217}}].

\bibitem{2021NatAs...5..465A}
A.U.~{Abeysekara}, A.~{Albert}, R.~{Alfaro}, C.~{Alvarez}, J.R.A.~{Camacho}, J.C.~{Arteaga-Vel{\'a}zquez} et~al., \emph{{HAWC observations of the acceleration of very-high-energy cosmic rays in the Cygnus Cocoon}}, \href{https://doi.org/10.1038/s41550-021-01318-y}{\emph{Nature Astronomy} {\bfseries 5} (2021) 465} [\href{https://arxiv.org/abs/2103.06820}{{\ttfamily 2103.06820}}].

\bibitem{1966ApJ...146..480J}
J.R.~{Jokipii}, \emph{{Cosmic-Ray Propagation. I. Charged Particles in a Random Magnetic Field}}, \href{https://doi.org/10.1086/148912}{\emph{\apj} {\bfseries 146} (1966) 480}.

\bibitem{2022FrASS...922100F}
K.~{Fang}, \emph{{Gamma-ray pulsar halos in the Galaxy}}, \href{https://doi.org/10.3389/fspas.2022.1022100}{\emph{Frontiers in Astronomy and Space Sciences} {\bfseries 9} (2022) 1022100} [\href{https://arxiv.org/abs/2209.13294}{{\ttfamily 2209.13294}}].

\bibitem{2021ApJ...923...53L}
A.~{Lazarian} and S.~{Xu}, \emph{{Diffusion of Cosmic Rays in MHD Turbulence with Magnetic Mirrors}}, \href{https://doi.org/10.3847/1538-4357/ac2de9}{\emph{\apj} {\bfseries 923} (2021) 53} [\href{https://arxiv.org/abs/2106.08362}{{\ttfamily 2106.08362}}].

\bibitem{2013Natur.497..466E}
G.~{Eyink}, E.~{Vishniac}, C.~{Lalescu}, H.~{Aluie}, K.~{Kanov}, K.~{B{\"u}rger} et~al., \emph{{Flux-freezing breakdown in high-conductivity magnetohydrodynamic turbulence}}, \href{https://doi.org/10.1038/nature12128}{\emph{\nat} {\bfseries 497} (2013) 466}.

\bibitem{2022MNRAS.512.2111H}
Y.~{Hu}, A.~{Lazarian} and S.~{Xu}, \emph{{Superdiffusion of cosmic rays in compressible magnetized turbulence}}, \href{https://doi.org/10.1093/mnras/stac319}{\emph{\mnras} {\bfseries 512} (2022) 2111} [\href{https://arxiv.org/abs/2111.15066}{{\ttfamily 2111.15066}}].

\bibitem{2023ApJ...959L...8Z}
C.~{Zhang} and S.~{Xu}, \emph{{Numerical Testing of Mirror Diffusion of Cosmic Rays}}, \href{https://doi.org/10.3847/2041-8213/ad0fe5}{\emph{\apjl} {\bfseries 959} (2023) L8} [\href{https://arxiv.org/abs/2311.18001}{{\ttfamily 2311.18001}}].

\bibitem{barreto2021}
L.~{Barreto-Mota}, E.M.~{de Gouveia Dal Pino}, B.~{Burkhart}, C.~{Melioli}, R.~{Santos-Lima} and L.H.S.~{Kadowaki}, \emph{{Magnetic field orientation in self-gravitating turbulent molecular clouds}}, \href{https://doi.org/10.1093/mnras/stab798}{\emph{\mnras} {\bfseries 503} (2021) 5425} [\href{https://arxiv.org/abs/2101.03246}{{\ttfamily 2101.03246}}].

\bibitem{2025ApJ...988..269B}
L.~{Barreto-Mota}, E.M.~{de Gouveia Dal Pino}, S.~{Xu} and A.~{Lazarian}, \emph{{Cosmic-Ray Diffusion in the Turbulent Interstellar Medium: Effects of Mirror Diffusion and Pitch-angle Scattering}}, \href{https://doi.org/10.3847/1538-4357/ade4c8}{\emph{\apj} {\bfseries 988} (2025) 269} [\href{https://arxiv.org/abs/2405.12146}{{\ttfamily 2405.12146}}].

\bibitem{Alves_Batista_2022}
R.~Alves~Batista, J.~Becker~Tjus, J.~Dörner, A.~Dundovic, B.~Eichmann, A.~Frie et~al., \emph{Crpropa 3.2 — an advanced framework for high-energy particle propagation in extragalactic and galactic spaces}, \href{https://doi.org/10.1088/1475-7516/2022/09/035}{\emph{Journal of Cosmology and Astroparticle Physics} {\bfseries 2022} (2022) 035}.

\bibitem{2024A&A...690A..94P}
B.~{Polak}, M.-M.~{Mac Low}, R.S.~{Klessen}, J.~{Wei Teh}, C.~{Cournoyer-Cloutier}, E.P.~{Andersson} et~al., \emph{{Massive star cluster formation: I. High star formation efficiency while resolving feedback of individual stars}}, \href{https://doi.org/10.1051/0004-6361/202348840}{\emph{\aap} {\bfseries 690} (2024) A94} [\href{https://arxiv.org/abs/2312.06509}{{\ttfamily 2312.06509}}].

\bibitem{2017ApJ...846...67P}
T.A.~{Porter}, G.~{J{\'o}hannesson} and I.V.~{Moskalenko}, \emph{{High-energy Gamma Rays from the Milky Way: Three-dimensional Spatial Models for the Cosmic-Ray and Radiation Field Densities in the Interstellar Medium}}, \href{https://doi.org/10.3847/1538-4357/aa844d}{\emph{\apj} {\bfseries 846} (2017) 67} [\href{https://arxiv.org/abs/1708.00816}{{\ttfamily 1708.00816}}].

\bibitem{2006PhRvD..74c4018K}
S.R.~{Kelner}, F.A.~{Aharonian} and V.V.~{Bugayov}, \emph{{Energy spectra of gamma rays, electrons, and neutrinos produced at proton-proton interactions in the very high energy regime}}, \href{https://doi.org/10.1103/PhysRevD.74.034018}{\emph{\prd} {\bfseries 74} (2006) 034018} [\href{https://arxiv.org/abs/astro-ph/0606058}{{\ttfamily astro-ph/0606058}}].

\end{thebibliography}\endgroup


\end{document}